\documentclass[twocolumn,showpacs,preprintnumbers,amsmath,amssymb,prb]{revtex4-1}
\usepackage{amsfonts}
\usepackage[english]{babel}
\usepackage[T1]{fontenc}
\usepackage{times}
\usepackage{mathrsfs}
\usepackage{graphicx}
\usepackage{dcolumn}
\usepackage{bm}
\usepackage[colorlinks,bookmarks=true,citecolor=blue,linkcolor=red,urlcolor=blue]{hyperref}
\usepackage[tight, FIGTOPCAP, normal, raggedright, nooneline]{subfigure}
\usepackage{hyperref}

\newcommand{\neff}{N'}

\begin{document}


\title{Topological equivalence of crystal and quasicrystal band structures}
\author{Kevin A.~Madsen}
\author{Emil J.~Bergholtz}
\author{Piet W.~Brouwer}
\affiliation{Dahlem Center for Complex Quantum Systems and Institut f\"ur Theoretische Physik, Freie Universit\"at Berlin, Arnimallee 14, 14195 Berlin, Germany}

\date{\today}

\begin{abstract}
A number of recent articles have reported the existence of topologically non-trivial states and associated end states in one-dimensional incommensurate lattice models that would usually only be expected in higher dimensions. Using an explicit construction, we here argue that the end states have precisely the same origin as their counterparts in commensurate models and that incommensurability does not in fact provide a meaningful connection to the topological classification of systems in higher dimensions. In particular, we show that it is possible to smoothly interpolate between states with commensurate and incommensurate modulation parameters without closing the band gap and without states crossing the band gap.
\end{abstract}

\maketitle

\section{Introduction}
A landmark result in recent years is that the band structure of (free-)fermion systems allows for an exhaustive topological classification of gapped phases.\cite{classification1,classificationkitaev,classification2} The rules of this classification are simple and natural: for a fixed chemical potential within a given band gap, two insulators are said to be topologically equivalent if and only if there exists an adiabatic deformation connecting the two insulators without closing the pertinent bulk gap, while if such a deformation does not exist, the insulators are said to be topologically distinct. The classification depends on fundamental symmetries, such as particle-hole or time-reversal symmetry, which apply to the insulators as well as to the deformations connecting them. Phases that are ``topologically nontrivial'' are characterized by gapless edge states which are robust in the sense that they cannot be removed without either closing the bulk band gap or breaking the underlying symmetry. 

In each spatial dimension there are five classes of topological insulators or superconductors depending on the underlying symmetry.\cite{classification1,classificationkitaev,classification2} Examples of topologically nontrivial phases are the quantum Hall effect and the quantum spin Hall effect in two dimensions,\cite{kane2005,bernevig2006b} as well as the recently discovered time-reversal-symmetric topological insulators in three dimensions\cite{moore2007,fu2007,roy2009,hsieh2008} (for reviews, see Refs.\ \onlinecite{koenig2008,hasan2010,qi2011}). In particular, if no symmetry is assumed, as in the case of the quantum Hall effect, nontrivial topological phases exist only if the number of spatial dimensions is even. 

The topological classification has been extended to families of insulators, in which the Hamiltonian $H$ depends periodically on a parameter $\phi$. The topological classification of such families of insulators then follows the classification of topological insulators in a higher dimension.\cite{fulga2012,roy2011} For example, whereas there are no topologically nontrivial insulators in one dimension if no symmetries are assumed, one-parameter families of one-dimensional band insulators share the same topological classification as the quantum Hall effect.\cite{laughlin1982,thouless} Similarly, one-parameter families of insulators with a time-reversal constraint share the same topological classification as the quantum spin-Hall effect in two dimensions.\cite{fu2006,meidan2010} In a topologically nontrivial family of insulators the existence of edge states is no longer guaranteed for all members of the family. Instead, it is only the existence of edge states for certain (usually odd) numbers of members of the insulator family that is topologically protected.

Recently the question was put forward as to whether quasicrystalline insulators are subject to the same topological classification as conventional crystalline band insulators.\cite{kraus} The question is interesting, because quasicrystals may be constructed by taking a cut through a standard crystal in a higher dimension,\cite{suck2002} thus motivating the prospect that quasicrystals could provide a way of probing higher-dimensional topological phases in lower-dimensional systems.\cite{kraus,lang,mei} In Ref.\ \onlinecite{kraus}, Kraus {\em et al.}\ investigated a model for a one-dimensional quasicrystalline insulator and showed that a family of such one-dimensional quasicrystals has a nontrivial topological classification, including the existence of topologically protected end states for some members of the family. (In one dimension, edge states are located at the two ends, which is why we refer to them as ``end states.'') This result is consistent with the general classification paradigm for families of one-dimensional insulators.\cite{thouless} However, Ref.\ \onlinecite{kraus} goes one step further and claims that each individual realization of such a quasicrystal already possesses the topological properties characteristic of a two-dimensional insulator. This claim, which is at odds with the standard classification of Refs.\ \onlinecite{classification1,classificationkitaev,classification2}, has been highlighted in the science media,\cite{quandt,iop} and explicitly acknowledged in several recent works.\cite{lang,mei,kraus2,xu,ganeshan,deng,segev,verbin,chen,ringel,grusdt,xuchen,langchen,satija} 

Although much of the recent work is interesting in its own right---e.g., the experiment reported in Ref. \onlinecite{kraus} describes a beautiful realization of a topological pump\cite{thouless}---we feel it is necessary to show by explicit construction that there is no topological distinction between one-dimensional insulating crystals and one-dimensional insulating quasicrystals. This means that the standard classification of topological phases is indeed correct and allows no exception for the case of quasicrystals. 

Our construction makes use of the Aubry-Andr\'e model,\cite{aubry} the same model for a one-dimensional quasicrystal as in Ref.\ \onlinecite{kraus}. Although this model does not describe a quasicrystal in the strict sense, the topological properties of the Aubry-Andr\'e model were shown to be the same as those of a true quasicrystal.\cite{kraus2} In Sec. \ref{sec_setup} we review the relevant features of this model, including the appearance of end states for certain parameters of the model. In Sec. \ref{sec_adiabatic} we go on to show that we can readily interpolate between the Aubry-Andr\'e model and a model for a standard crystalline insulator in one dimension without closing the bulk gap. In fact, we are able to provide a stronger result, namely, that it is possible to interpolate between these models even without end states crossing the gap. In Sec. \ref{sec_movie} we consider a family of one-dimensional insulators and show that a family of quasicrystalline insulators can be adiabatically mapped to a family of standard crystalline insulators, again without states closing the excitation gap. Hence, while this is not true for the individual members, a topological index can meaningfully be assigned to a family of one-dimensional Hamiltonians, but this index does not distinguish between crystalline and quasicrystalline insulators. We close by summarizing our conclusions in Sec. \ref{conclusions}.

\begin{figure}
\centerline{\includegraphics[width=\linewidth]{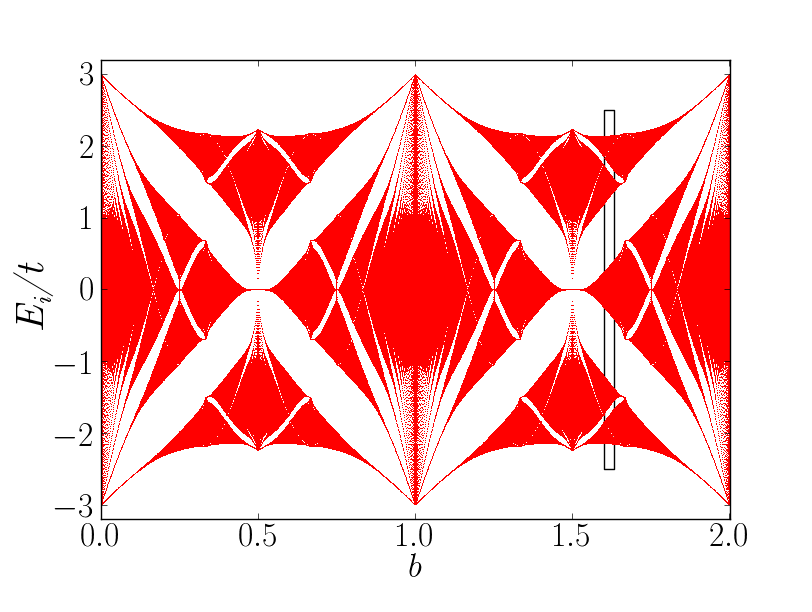}}
\caption{(Color online) The full spectrum of the Aubry-Andr\'e model (\ref{eq_aamodel}) with periodic boundary conditions for $b$ from $0$ to $2$. System size $N=1000$ and the amplitude of the modulating potential $\lambda/t=1$. Note that the spectrum is periodic for $b \rightarrow b+1$. With periodic boundary conditions this spectrum  has no end states. The black rectangle marks the interval of $b$ that is investigated in detail in the remainder of the article.}
\label{fig_pbc_butterfly}
\end{figure}

\section{Aubry-Andr\'e model}
\label{sec_setup}

The Audry-Andr\'e model describes electrons hopping on a one-dimensional chain, with a spatially modulated on-site potential, such that the period of the potential is incommensurate with the lattice period,\cite{aubry}
\begin{align}
	\label{eq_aamodel}
	H(\phi) \psi_n = t (\psi_{n+1} + \psi_{n-1}) + \lambda \cos(2 \pi b n + \phi) \psi_n .
\end{align}
Here $\psi_n$ is the wave function at site $n=0,1,\ldots ,N-1$, $t$ is the nearest-neighbor hopping amplitude, $\lambda$ is the amplitude of the on-site potential, and $b$ is its inverse period. The parameter $\phi$ describes the phase of the potential with respect to the lattice at position $n=0$. The family of one-dimensional models $\{H(\phi)|0<\phi\leq 2\pi\}$ defines a model in two dimensions (and may as such be topological). When the inverse period $b$ is irrational, the potential is incommensurate with the lattice, and the resulting state serves as a model for a quasicrystal. For a rational inverse period $b$ the model (\ref{eq_aamodel}) describes a conventional crystalline band insulator.

The Audry-Andr\'e model has been well studied in the literature. We here summarize the features most important for our subsequent analysis. 

\begin{figure}
\subfigure{\includegraphics[width=0.8\linewidth]{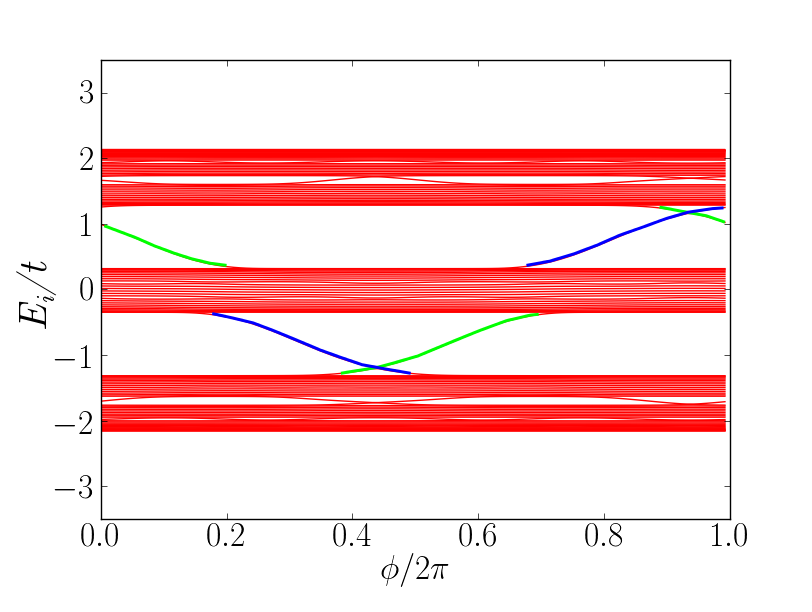}\label{fig_phiscan1}}
\subfigure{\includegraphics[width=0.8\linewidth]{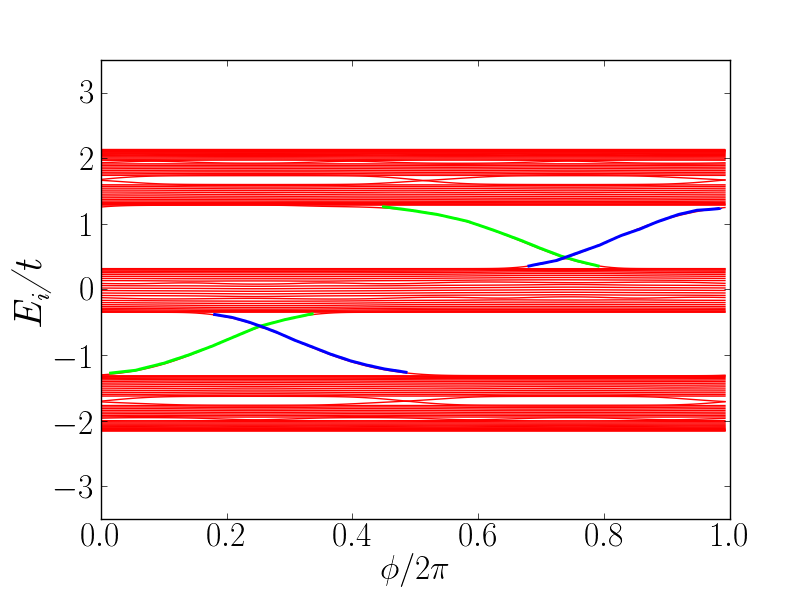}\label{fig_phiscan2}}
\caption{(Color online) Spectra of a finite-length Aubry-Andr\'e chain as a function of $\phi$. The chain length is $N=108$. The other model parameters are set at $\lambda/t=1$ and $b=(\sqrt5+1)/2$. In the lower panel a potential of the form (\ref{eq_wallpotential}) with $m/t = 10$ and $\neff = 106$ has been added to the last two sites near the right end of the chain. End states located on the left are colored blue, and end states on the right are green (see Fig. \ref{fig_example_endstate} for an example of such an end state). Only the positions of the right end states (green) are affected by the modified end potential. \label{fig_phiscan}}
\end{figure}

The (bulk) energy spectrum of Eq.\ (\ref{eq_aamodel}) shows a self-similar structure as a function of the inverse period $b$, resembling the Hofstadter butterfly;\cite{hofstadter} see Fig. \ref{fig_pbc_butterfly}. For a fixed value of $b$ there are both large and small gaps in the spectrum, only a small number of which can be detected in the limited resolution of Fig.\ \ref{fig_pbc_butterfly}. If $b$ is irrational, the gap structure is fractal. The single-particle states are localized for $|\lambda/t|>2$; they are extended if $|\lambda/t|<2$.\cite{aubry} In either case, the system is an insulator if the Fermi energy lies on one of the gaps.

\begin{figure}
\includegraphics[width=0.8\linewidth]{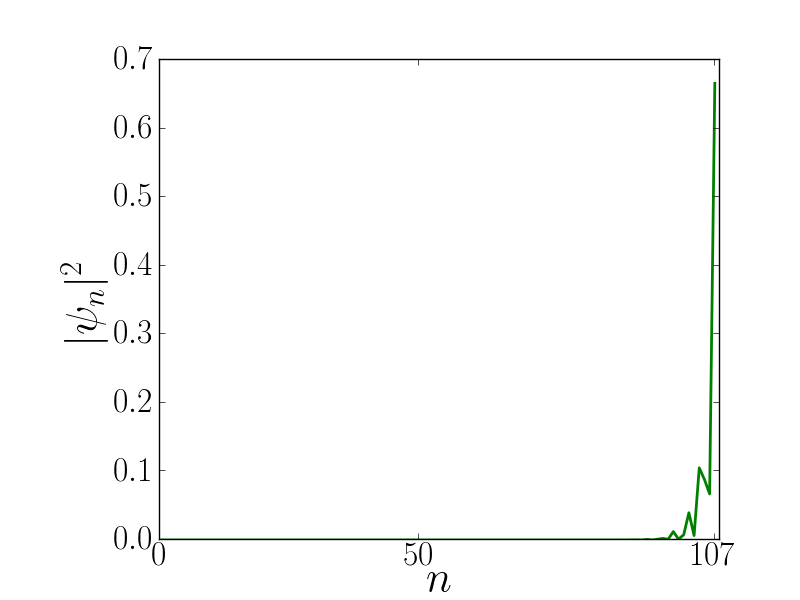}
\caption{(Color online) Density $|\psi_n|^2$ as a function of the lattice position $n$ for an end state localized near the right end of the chain. The model parameters have been chosen as in the top panel of Fig.\ \ref{fig_phiscan}. The energy eigenvalue is $E/t = 1.04$, placing it in the upper large gap at $\phi=0$. \label{fig_example_endstate}}
\end{figure}

Keeping $b$ fixed and varying $\phi$, now for a lattice of finite length $N$ and with open boundary conditions, one obtains a spectrum as shown in Fig. \ref{fig_phiscan}. The spectrum shows the same gap structure as in Fig.\ \ref{fig_pbc_butterfly}, but it also shows end states crossing the gaps at certain values of $\phi$. The color coding in the figure is such that end states near the left end (near $n=0$) are colored blue, whereas end states near the right end (near $n=N-1$) are colored green. An example of the wave function of such an end state is shown in Fig.\ \ref{fig_example_endstate}.

\begin{figure}
\centerline{\includegraphics[width=0.8\linewidth]{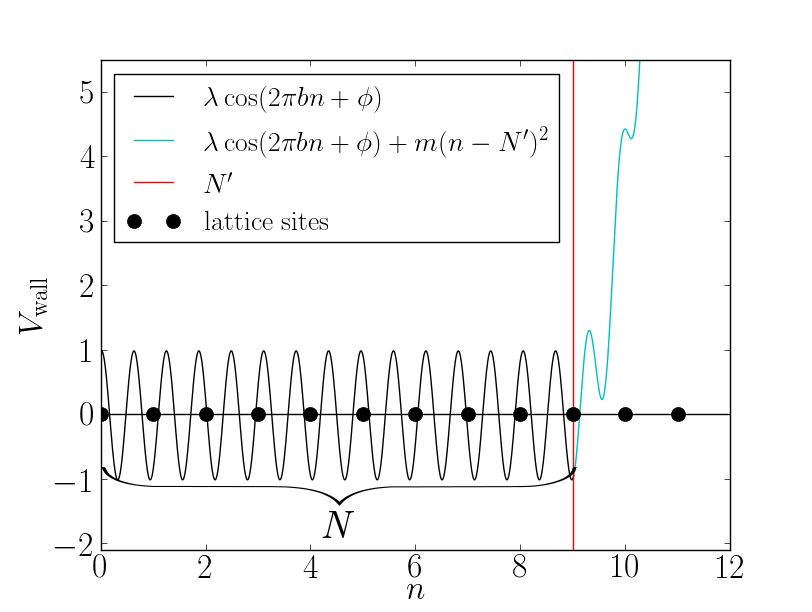}}
\caption{(Color online) Total on-site potential in the Aubry-Andr\'e model (\ref{eq_aamodel}) with the additional termination potential (\ref{eq_wallpotential}) at the chain's right end. Parameters have been set at $b=(\sqrt5+1)/2$, $\lambda/t=1$, $\phi=0$, $m/t=4$, $N=12$ and $\neff =9$.
\label{fig_potential}}
\end{figure}

The existence of the end states for the $\phi$-dependent family of insulators is topologically protected,\cite{kraus} but the precise value of $\phi$ for which the end states occur depends on the details of the termination of the lattice. Figure \ref{fig_phiscan} (bottom) shows the $\phi$-dependent spectrum for the case that an additional potential of the form 
\begin{align}
	\label{eq_wallpotential}
	V_{\rm wall} = \left\{ 
	\begin{array}{cl}
		m (n-\neff )^2  & \mbox{if } n>\neff \ , \\
		0  & \mbox{if } n \leq \neff  \ ,
	\end{array}
	\right.
\end{align}
with $\neff  = N-2$, has been added to the last two sites at the right end of the chain. The combined potential is sketched in Fig. \ref{fig_potential} for a lattice with $N=12$ and $\neff  = 9$. Effectively, the potential $V_{\rm wall}$ truncates the Aubry-Andr\'e chain at length $\neff $. Addition of the potential $V_{\rm wall}$---i.e., effectively shortening the chain---shifts the end states located on the right (marked green in Fig. \ref{fig_phiscan}) to a different value of $\phi$ inside the gap, as shown in Fig.\ \ref{fig_phiscan} (bottom). The end states located on the left end of the chain remain unaffected by the modified potential because $V_{\rm wall}$ affects the right end of the chain only.

The strong dependence on termination indicates that the end states are fragile in a one-dimensional model at fixed $\phi$, since any end state can readily be removed from the Fermi energy by modifying the potential near the chain ends. The absence of topological protection for the end states at fixed $\phi$ was noticed in Ref.\ \onlinecite{kraus}, yet it did not lead to the conclusion that the Aubry-Andr\'e model at fixed $\phi$ is topologically trivial. Below we will explicitly demonstrate that topological properties of the model are the same for irrational and rational $b$ by showing that one can always choose the termination of the model at fixed $\phi$ in such a way that end states are eliminated for a finite interval of $b$. For rational $b$ the model (\ref{eq_aamodel}) is known to be topologically trivial.

\section{Adiabatic continuity between rational and irrational inverse periods}
\label{sec_adiabatic}

\begin{figure}
\subfigure{\includegraphics[width=0.8\linewidth]{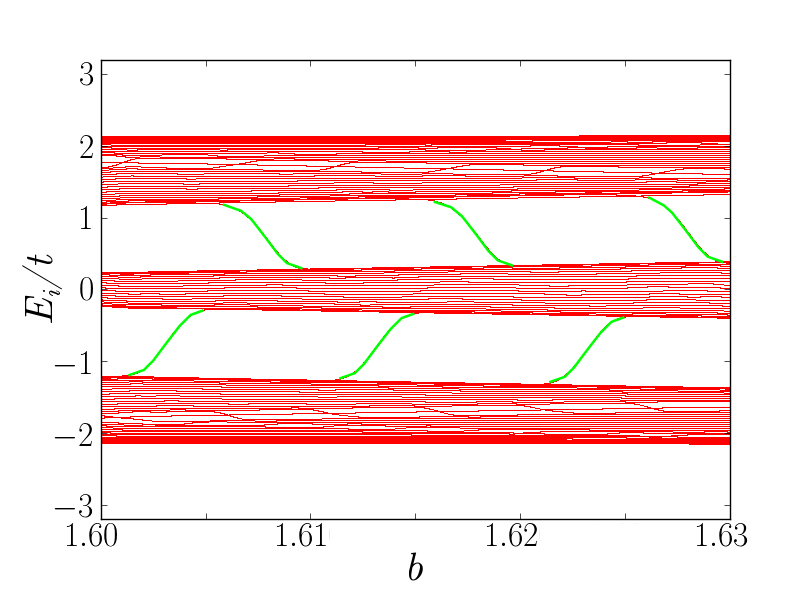} \label{fig_bscan_nopot1}}
\subfigure{\includegraphics[width=0.8\linewidth]{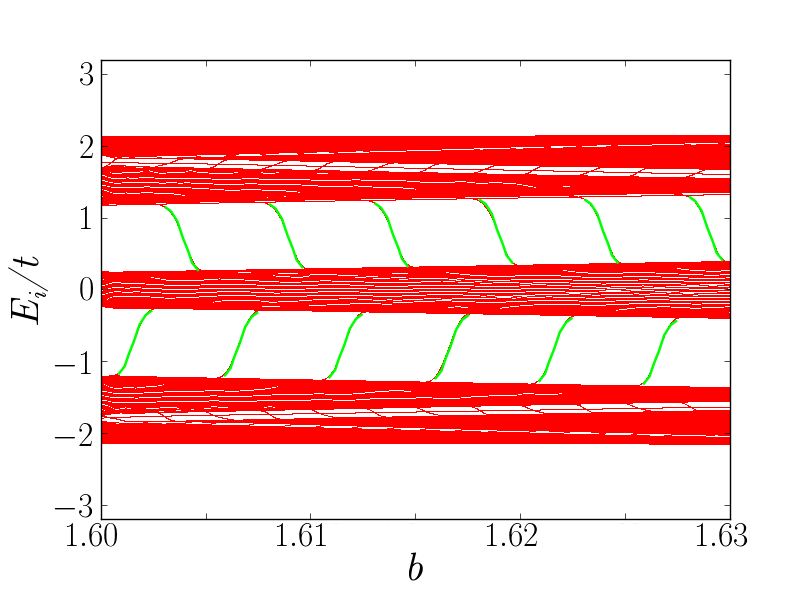} \label{fig_bscan_nopot2}}
\caption{(Color online) Spectra of the Aubry-Andr\'e model with $\lambda/t=1$ and $\phi=0$ as a function of the inverse period $b$. In the top panel $N=108$; in the bottom panel $N=216$. A potential of the form (\ref{eq_wallpotential}) with $m/t=10$ has been added at the right end of the chain, with $\neff  = 99$ (top) and $\neff =199$ (bottom), effectively limiting the number of lattice sites to $100$ and $200$, respectively. End states located on the right are colored green. 
\label{fig_bscan}}
\end{figure}

Figure \ref{fig_bscan} shows the spectrum for $1.60 < b < 1.63$ and two different system sizes $N$. It can be seen that the four largest spectral gaps do not close in this interval. We will focus our attention on the largest and second to largest gaps in the lower half of the spectrum. 

The figure clearly shows states that cross to the center band (centered around $E/t = 0$) and to the lowest band. (We also expect to see such crossings for the smaller gaps, but these can not be distinguished for the finite-size systems in Fig.\ \ref{fig_bscan}.) The origin of these gap-crossing states is particle-number conservation: since the number of states inside each band changes with $b$, particle-number conservation requires that states must cross between the bands. The number of such crossings is proportional to the chain length $N$ and to the amount by which the inverse period $b$ is changed. Since in our formulation (\ref{eq_aamodel}) of the Aubry-Andr\'e model the phase of the potential is fixed near the chain's left end (at $n=0$) upon varying $b$, and since the bulk remains insulating throughout, these gap crossings must take place as in-gap end states at the right end of the chain. This suggests that the gap-crossing states can be eliminated by suitably adapting the chain length near the right end. We will now demonstrate that this is indeed the case.

\begin{figure}
\subfigure{\includegraphics[width=0.8\linewidth]{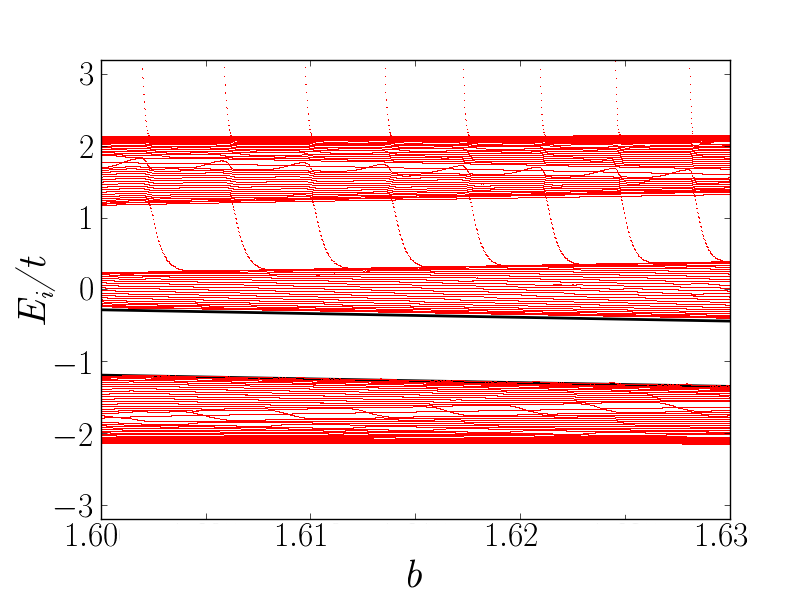} \label{fig_bscan_108}}
\subfigure{\includegraphics[width=0.8\linewidth]{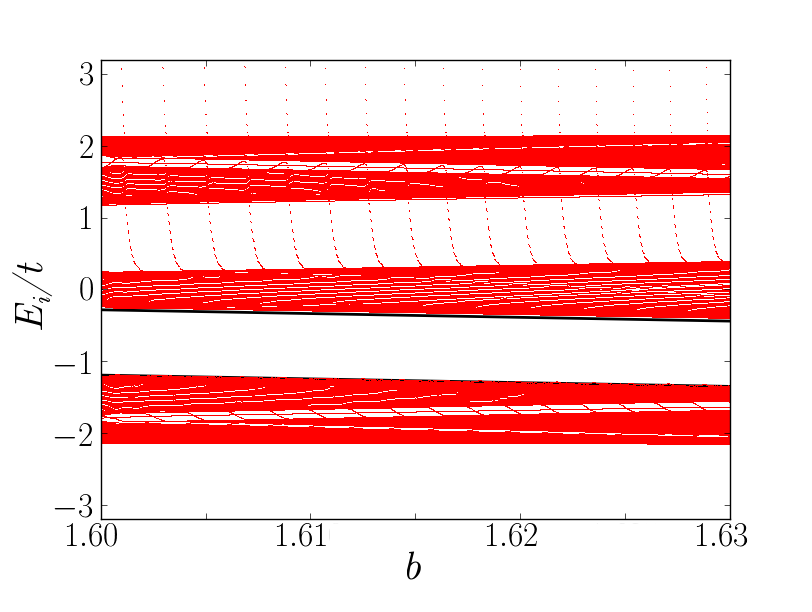} \label{fig_bscan_216}}
\caption{(Color online) Spectrum of the Aubry-Andr\'e model as a function of the inverse period $b$ with parameters $\lambda/t=1$ and $\phi=0$. At the right end, the chain is terminated with a potential of the form (\ref{eq_wallpotential}) with $m/t=10$. The effective length $\neff $ is adjusted following the procedure described in the main text, such that $\neff$ increases from $\neff =99$ at $b=1.60$ to $\neff  = 106.9$ at $b=1.63$ (top panel) as can be seen in detail in Fig. \ref{fig_Nprime_over_b}, and from $\neff =199$ at $b=1.60$ to $\neff  = 214.9125$ at $b=1.63$ (bottom panel). The total chain length is kept at $N=108$ and $N=216$, respectively. The black lines mark the thresholds of the algorithm described above.
\label{fig_bscan_noedge}}
\end{figure}

To this end we include the potential (\ref{eq_wallpotential}) into the model, choose the total chain length $N$ sufficiently larger than the effective chain length $\neff $, and determine the effective chain length $\neff $ by a simple algorithm: 
as $b$ is varied, we calculate the spectrum for each value of $b$ and monitor eigenvalues that lie inside the gap between two fixed thresholds. (The thresholds are shown as black lines in Fig. \ref{fig_bscan_noedge}.) If an eigenvalue is about to enter into the gap from below, we increase $\neff $ by a small amount and recalculate the spectrum with the new value of $\neff $. Similarly, if an eigenvalue is about to cross into the gap from above, we decrease $\neff $ by a small amount. We repeat this process for each $b$. As long as $N$ is sufficiently larger than the effective chain length $\neff $ throughout the $b$ interval of interest, its precise value does not affect the spectrum below and in the vicinity of the gap.

\begin{figure}
\includegraphics[width=0.8\linewidth]{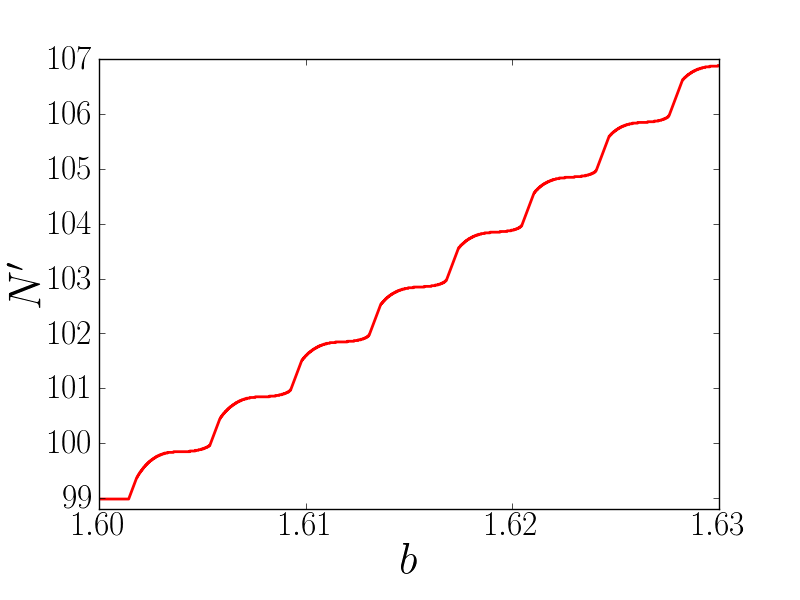}
\caption{(Color online) The effective chain length $\neff$ as a function of the inverse period $b$, for the data set shown in the top panel of Fig.\ \ref{fig_bscan_noedge}. \label{fig_Nprime_over_b}}
\end{figure}

Figure \ref{fig_bscan_noedge} shows how the end states vanish from the gap as a result of smoothly adjusting the effective chain length $\neff$ as a function of $b$, following the procedure described above. The value of the effective chain length $\neff$ is shown in Fig. \ref{fig_Nprime_over_b}. 
Note that the number of states that cross other gaps than the one of interest (in particular, the higher-energy gap near $E/t = 1$ in the figure), has increased as a result of this procedure. Also note that the chain length has to increase by an amount proportional to the chain length in order to preserve the number of states below the Fermi level, consistent with the expectation that the number of states inside each band is proportional to the chain length.

\begin{figure}
\subfigure{\includegraphics[width=0.8\linewidth]{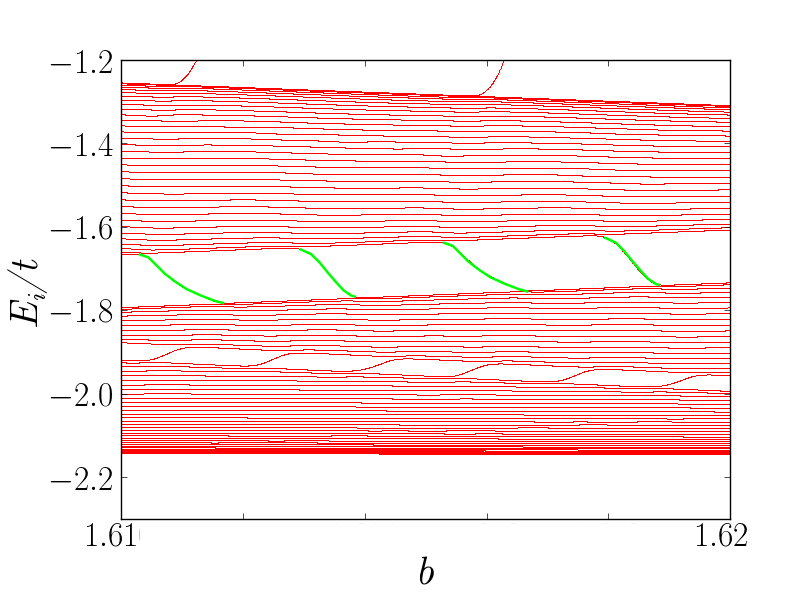} \label{fig_bscan_smallgap_nopot}}
\subfigure{\includegraphics[width=0.8\linewidth]{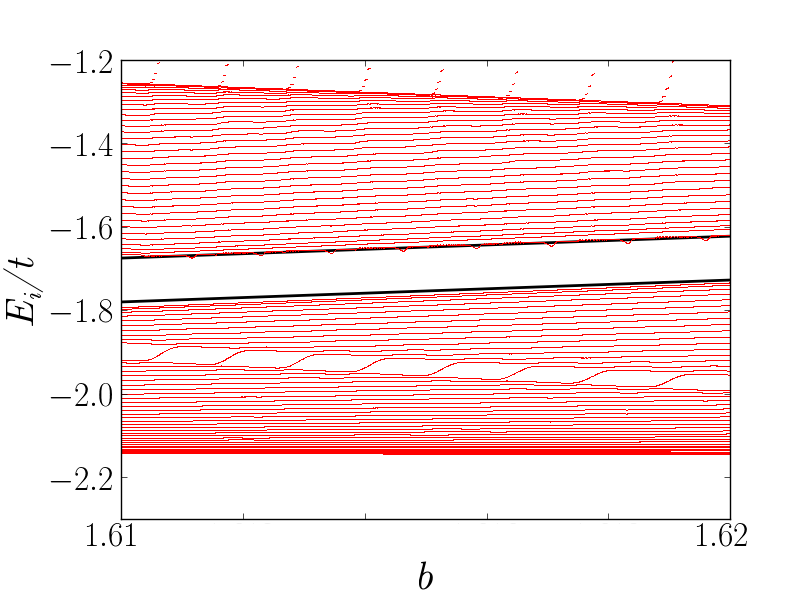} \label{fig_bscan_smallgap_noedge}}
\caption{(Color online) 
Spectrum of the Aubry-Andr\'e model near the second-largest gap in the lower half of the spectrum. Parameters are $\lambda/t = 1$, $\phi=2$ and $m/t=10$. The upper panel shows the spectrum at fixed effective length ($\neff  = 199$). In the lower panel the effective length is adjusted following the procedure described in the main text. The effective length $\neff $ decreases from $\neff  = 199$ at $b=1.61$ to $\neff  = 183.08$ at $b=1.62$. End states located near the right end are shown green. The black lines mark the thresholds of the algorithm described above.
\label{fig_bscan_smallgap}}
\end{figure}

The procedure can also be applied to the second-largest gap in the lower half of the spectrum, see Fig. \ref{fig_bscan_smallgap}. Now the effect of increasing $b$ on the spectrum at fixed length is inverted: upon increasing $b$, the end states are crossing into the gap from the higher band to the lower band, so we must effectively reduce the chain length upon increasing $b$. 

Although our numerical calculations take place at finite system size, we observe that the gaps remain finite and system-size independent throughout the deformation [compare Figs.\ \ref{fig_bscan_noedge} (top) and \ref{fig_bscan_noedge} (bottom)]. Thus, our results clearly show that two Aubry-Andr\'e chains with different inverse periods $b$ can be smoothly deformed into one another without closing the bulk gap. Moreover, if we allow the chain length to change during the deformation, the deformation can be performed in such a way that end states remain completely absent when interpolating between different inverse periods. Since the deformation covers both rational and irrational inverse periods $b$, we conclude that there is no topological difference between the Aubry-Andr\'e model with irrational $b$, which serves as a model for a quasicrystal, and the model at rational $b$, which describes a conventional one-dimensional band insulator.

\section{Adiabatic continuity for a one-parameter family of insulators}
\label{sec_movie}

In the previous section, we have shown that the Aubry-Andr\'e model at fixed phase $\phi$ with irrational inverse period $b$ can be adiabatically deformed into a model with a rational inverse period, where no states cross the Fermi energy if the chain length is suitably adjusted. We now show that such a deformation can be achieved for the complete family of one-dimensional models $\{H(\phi)|0<\phi\leq 2\pi\}$. If $\phi$ is interpreted as time, $\{H(\phi)|0<\phi\leq 2\pi\}$ describes the pumping in a one-dimensional time-periodic system,\cite{thouless} while if $\phi$ is interpreted as the transverse momentum it relates to the Hofstadter model of two-dimensional charged particles in a magnetic field. Like these systems, the family of models $\{H(\phi)|0<\phi\leq 2\pi\}$ can be topologically classified and assigned an integer Chern number.\cite{thouless1982} 

\begin{figure}
\includegraphics[width=0.42\linewidth]{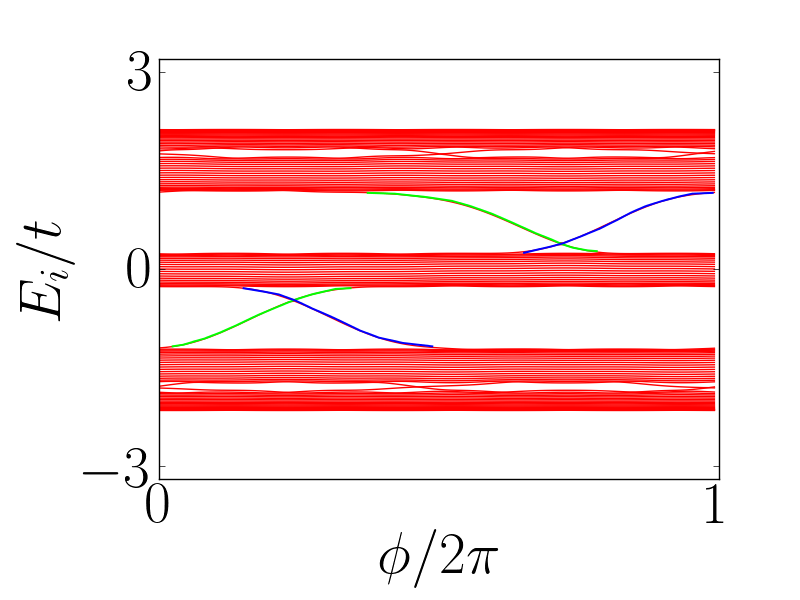}
\includegraphics[width=0.42\linewidth]{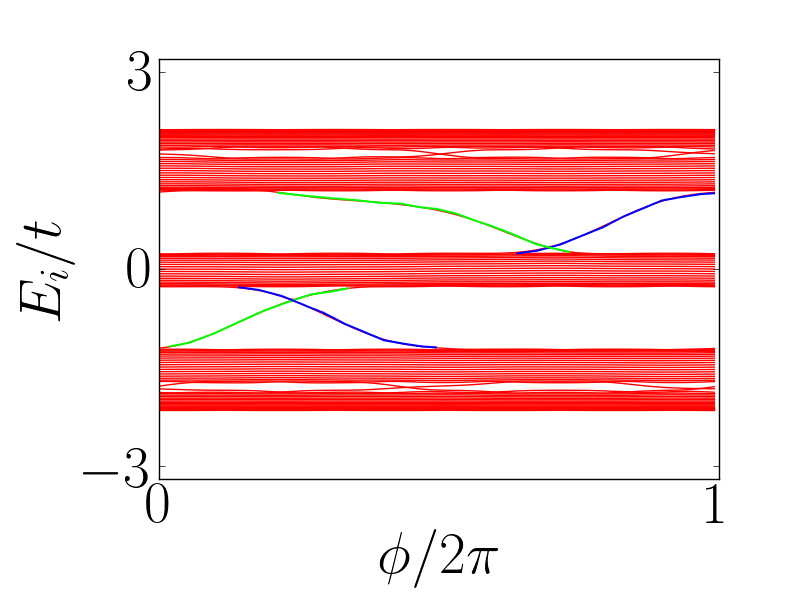}
\includegraphics[width=0.42\linewidth]{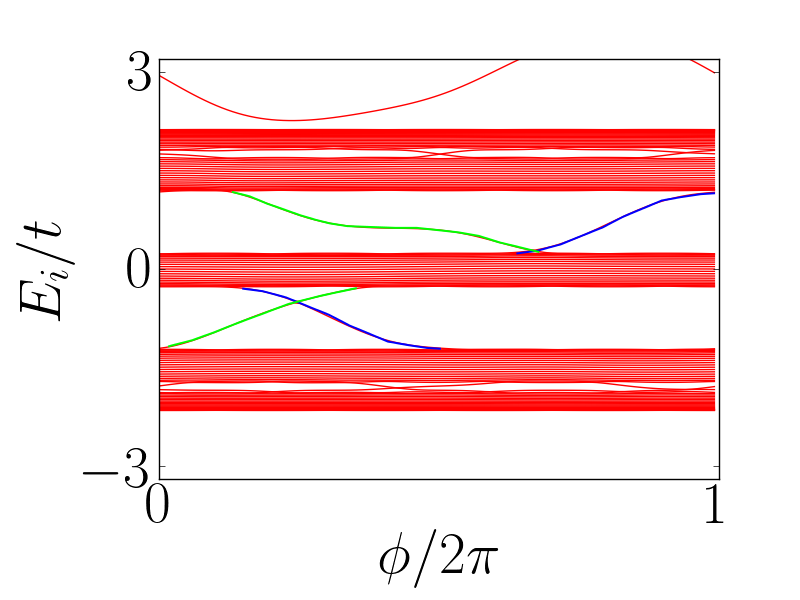}
\includegraphics[width=0.42\linewidth]{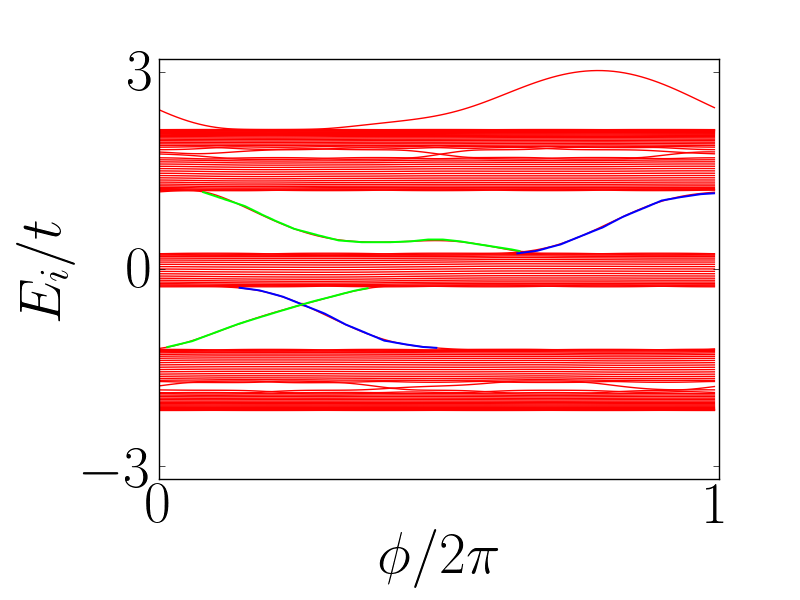}
\includegraphics[width=0.42\linewidth]{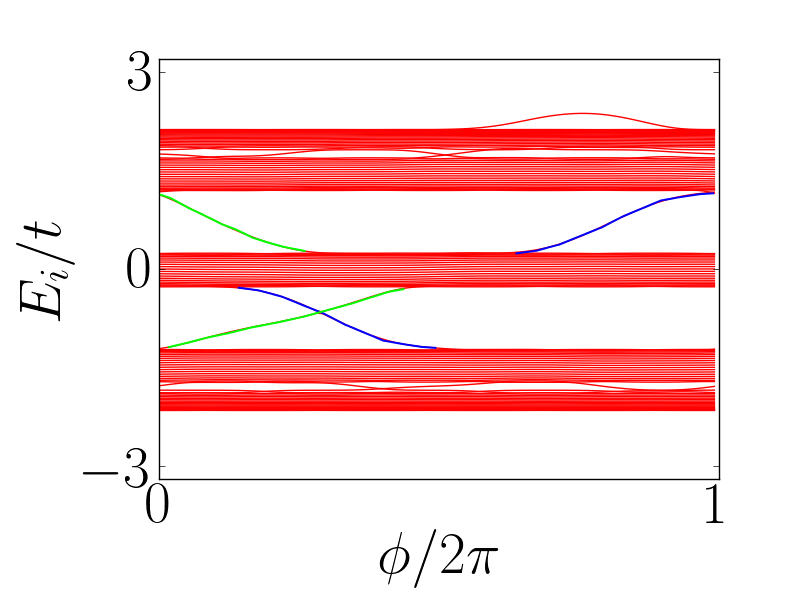}
\includegraphics[width=0.42\linewidth]{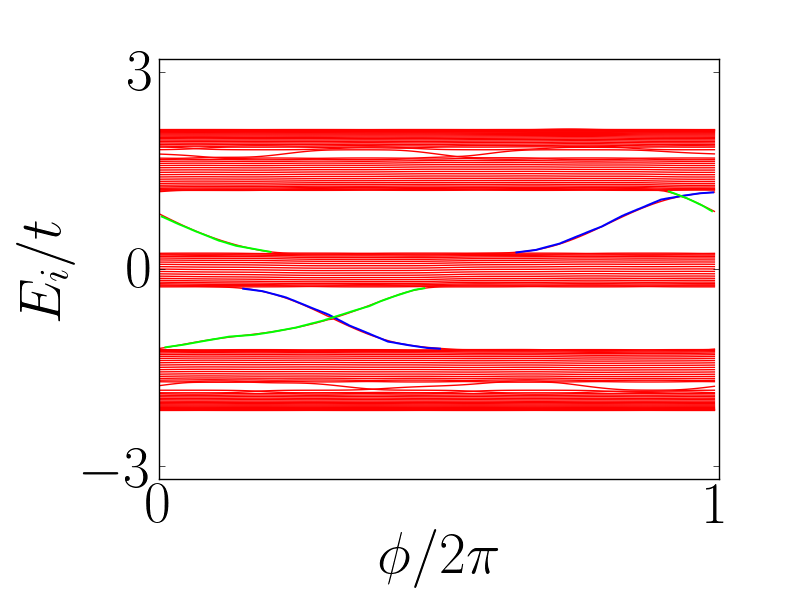}
\includegraphics[width=0.42\linewidth]{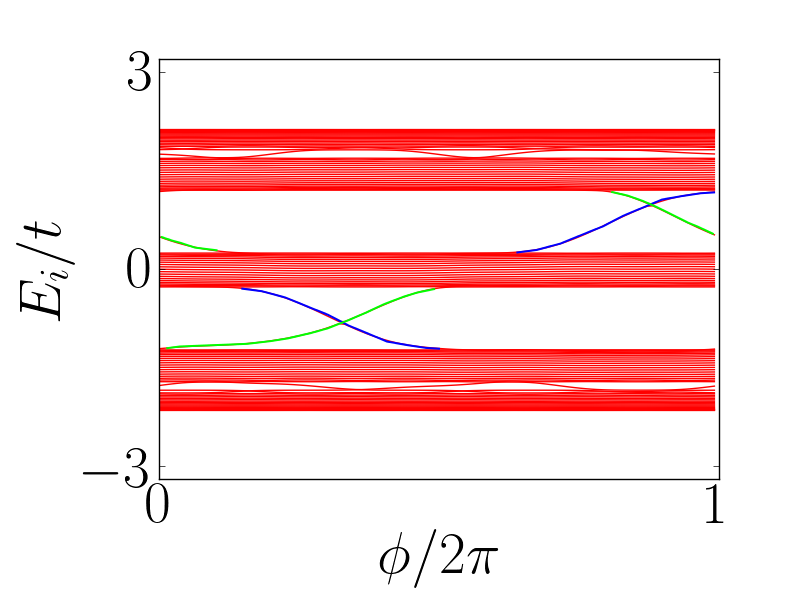}
\includegraphics[width=0.42\linewidth]{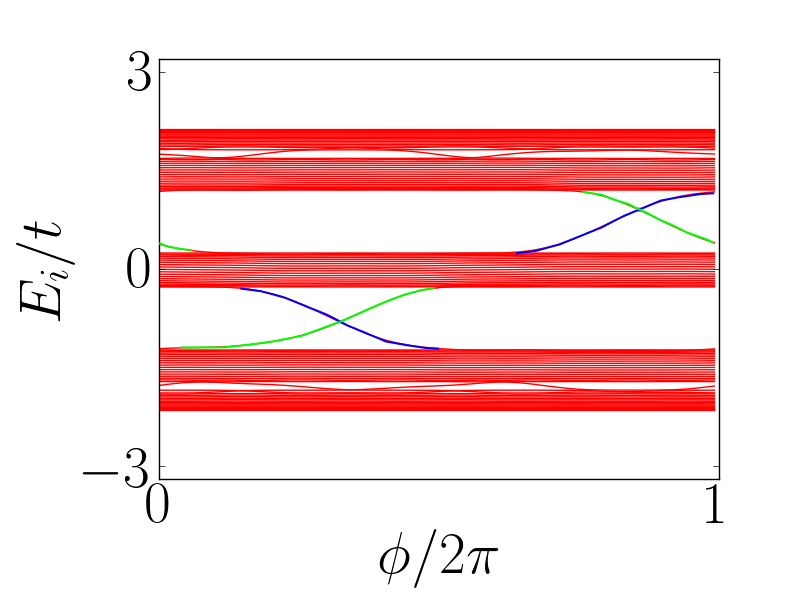}
\caption{(Color online) Spectra of the Aubry-Andr\'e model as a function of $\phi$, for $\lambda/t=1$ and for different values of the inverse period $b$, 
$b=1.601\,487\,5$, 
$1.601\,737\,5$, 
$1.601\,987\,5$, 
$1.602\,112\,5$, 
$1.602\,362\,5$, 
$1.602\,612\,5$, 
$1.602\,987\,5$, 
$1.603\,237\,5$ from top left to bottom right. A potential of the form (\ref{eq_wallpotential}) with $m/t=10$ is included and the effective chain length $\neff $ is determined by application of the procedure of Sec.\ \ref{sec_adiabatic} to the case $\phi=0$ and ranges from $\neff  = 99.1$ (at $b=1.6014875$) to $\neff  = 99.8375$ (at $b=1.6032375$). End states located on the left end are colored blue, states at the right are shown green.
\label{fig_movie}}
\end{figure}

Applying the procedure as in the previous section, adjusting the effective chain length $\neff $ such that no end states cross the large gap near $E/t=-1$ at $\phi=0$, we have studied the evolution of the entire $\phi$-dependent spectrum upon varying $b$.  In Fig. \ref{fig_movie} we show a selection of spectra for different values of $b$. We conclude that the topological class of the family of insulators $H(\phi)$ is independent of $b$ for the range of inverse periods shown in the figure and for the Fermi energy inside the large gap near $E/t = -1$. Our results can easily be extended to other intervals for the inverse period $b$ and to Fermi energies located inside other gaps, although the detailed dependence of the effective chain length $\neff $ on $b$ will be different in each case.

Families of one-dimensional band insulators are topologically characterized using the Chern number.\cite{thouless1982,thouless} The Chern number counts the number of end states at each end that cross the Fermi level as a function of the periodic parameter $\phi$. (The counting occurs with sign depending on whether the end states approach the Fermi energy from below or from above upon increasing $\phi$.) Our result explicitly demonstrates that the entire family of one-dimensional models $\{H(\phi)|0<\phi\leq 2\pi\}$ belongs to the same topological phase for all $b$, as long as the Fermi energy remains inside a gap. Hence, there are no differences between crystalline band insulators (corresponding to rational inverse period $b$) and quasicrystalline insulators (corresponding to irrational inverse period $b$ in the Aubry-Andr\'e model). For the case shown in Fig.\ \ref{fig_movie} the Chern numbers are $C=1$ ($C=2$) for the larger (smaller) gap, independent of $b$, corresponding to one (two) end state(s) at each end per $\phi$-cycle.

\section{Conclusion}\label{conclusions}

In this article we have investigated the adiabatic transformation of the Aubry-Andr\'e model with irrational inverse period $b$ to a model with rational inverse period $b$. If the transformation occurs at a fixed chain length, the deformation can take place without closing a bulk excitation gap, provided the change of the inverse period $b$ is small enough. (The spectrum at irrational $b$ has a fractal gap structure, and the smaller gaps persist only over a small range of inverse periods $b$.) In principle, the absence of a bulk gap closing is enough to argue that the cases of rational $b$ and irrational $b$ are topologically equivalent.

For deformations that take place at fixed chain length $N$, discrete end states cross the bulk gaps upon changing the inverse period $b$. The origin of these gap crossings is particle-number conservation: the number of states available in the ``occupied bands'' changes as a function of the inverse period $b$, so that individual states are ``forced'' through the gap upon changing $b$. These end states can be eliminated if the chain length is adjusted while varying $b$. We have implemented a procedure in which the effective chain length can be adapted continuously, and have shown that this procedure leads to a deformation between different values of $b$ in which no states whatsoever cross the Fermi energy. We have extended these results to a one-dimensional family of insulators.

Since conventional band insulators in one dimension (with a periodic potential) are known to be topologically trivial, we conclude that the Aubry-Andr\'e model (with irrational inverse period) is topologically trivial, too. The Aubry-Andr\'e model has been shown to be in the same topological class as one-dimensional quasicrystals,\cite{kraus2} so that we may extend this conclusion to quasicrystalline insulators in one dimension. Hence, we confirm the validity of the general classification scheme of Refs.\ \onlinecite{classification1,classificationkitaev,classification2} to quasicrystalline insulators in one dimension.
Similarly, since one-parameter families of band insulators in one dimension are described by an integer invariant, the Chern number,\cite{thouless} we conclude that the same applies to families of one-dimensional quasicrystalline insulators.

While the examples provided here have concentrated on one-dimensional (quasi-)crystals, we see no reason why they would not, \textit{mutatis mutandis}, apply also to higher-dimensional generalizations of the scenario outlined in Ref. \onlinecite{kraus}. This applies, e.g., to the claim that four-dimensional quantum Hall effects may be observed in two-dimensional quasi-crystals.\cite{ringel}

\bigskip

{\it Acknowledgements.}~
We gratefully acknowledge discussions with Tobias Micklitz, Bj\"orn Sbierski, and Oded Zilberberg.
This work is supported by the Alexander von Humboldt Foundation in the framework of the Alexander von Humboldt Professorship, endowed by the Federal Ministry of Education and Research, and by the Emmy Noether program (Grant No. BE 5233/1-1) of the German Research Foundation (DFG). 

\bibliography{bibliography}

\end{document}